\documentclass{JINST}

\usepackage{graphicx}
\usepackage{booktabs}
\usepackage{textcmds}

\title{Energy deposition studies for the Neutrino Factory target station}

\author{John J Back$^a$ \\
\llap{$^a$} Department of Physics, University of Warwick,\\ 
Gibbet Hill Road, Coventry, CV4 7AL, UK\\
E-mail: \email{J.J.Back@warwick.ac.uk}}

\abstract
{A study of the energy deposition in the Neutrino Factory baseline target station
is presented. FLUKA simulations are used to study how
much of the 4\,MW proton beam power is distributed within the
target station, and these results are compared with MARS
simulations. About 10\% of the beam power is deposited in the mercury jet target,
with more than 2\,MW absorbed by the tungsten-carbide shielding.
Heating in the superconducting coils,
that generate most of the magnetic field in the target region, 
can be substantially reduced by increasing the surrounding shielding.
Estimates of radiation damage are also given for different sections
of the target geometry.}


\begin{document}

\section{Introduction}

\begin{figure}[hbt]
\centering
\includegraphics*[width=0.8\textwidth]{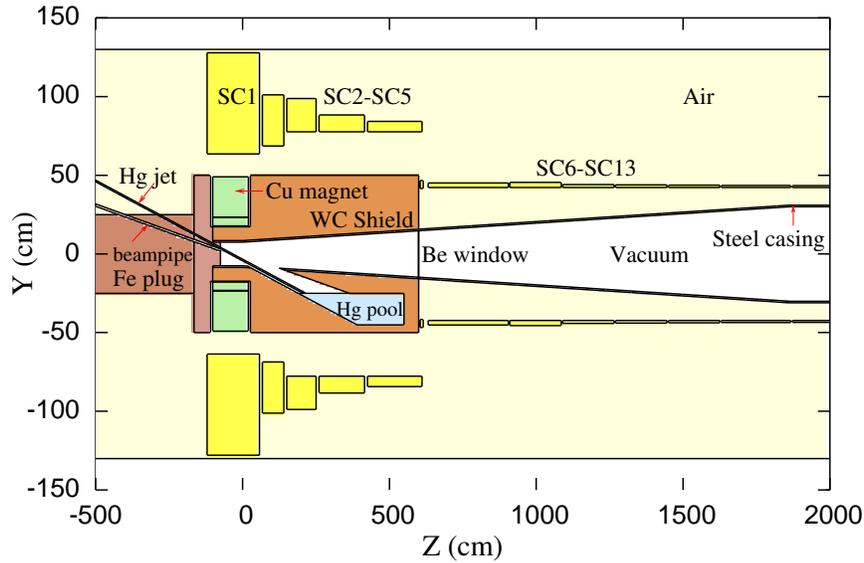}
\caption{Schematic of the Neutrino Factory target system. 
The superconducting magnets are labelled SC$n$, where $n$ = 1 to 13.}
\label{fig:standardGeometry}
\end{figure}

The current baseline option for the Neutrino Factory is to use
a 4\,MW proton beam interacting with a free-flowing mercury jet
to create an intense muon beam. 
The MERIT experiment has shown a proof-of-principle demonstration of 
a high intensity liquid mercury jet target~\cite{ref:MERIT}.
Figure~\ref{fig:standardGeometry} shows a schematic of the Neutrino 
Factory target station. The interaction of the proton beam with
the mercury jet creates low-energy pions that are captured
by the high field ($\sim20$\,T) solenoid and transported
into a decay channel where the muons resulting from pion decay are
collected and stored until they decay to neutrinos.

Despite the success of the MERIT experiment, further work was necessary
to investigate how the 4\,MW beam power is distributed inside the target
station so that adequate shielding can be put in place to 
protect vulnerable systems such as the superconducting magnets. 
This paper presents a study of the power deposition and radiation environment
of the Neutrino Factory target station using FLUKA~\cite{ref:FLUKACode} 
simulations. Also shown are comparisons with results from
MARS~\cite{ref:MARSCode,ref:MARSEDep} simulations.

\section{Simulation Parameters}

As a starting point, the Neutrino Factory mercury jet target station
geometry is based on the Study 2a configuration~\cite{ref:Study2a}, as shown 
in figure~\ref{fig:standardGeometry}, with the appropriate 20\,T field map 
based on the dimensions and currents in the normal conducting copper ($\sim 6$\,T) 
and superconducting coils ($\sim 14$\,T).
Variations to the geometry are made to investigate whether the energy deposition in the 
superconducting magnets can be reduced.
The proton beam has a Gaussian profile with a root mean square radius of 1.2\,mm,
and is tilted by approximately 100\,mrad with respect to the magnetic ($z)$ axis.
The kinetic energy of the proton beam is nominally set to 8\,GeV, while the 
mercury jet is modelled as a simple cylinder with a radius of 4\,mm. 
The angle between the mercury jet and the proton beam at their intersection
($z = -37.5$\,cm) is 27\,mrad, which is chosen to optimise the production of useful
low-energy pions from the target~\cite{ref:MARSPion}. 

\section{Study 2a geometry}

\begin{figure}[hbt]
\centering
\includegraphics*[width=0.9\textwidth]{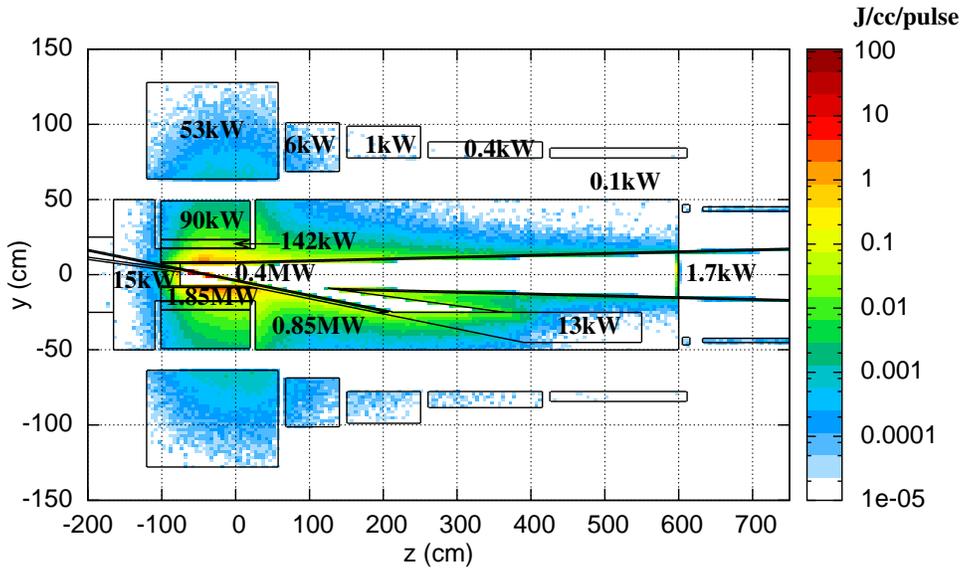}
\caption{Distribution of the deposited energy density (J/cc) per beam pulse
(50\,Hz repetition rate) in the Study 2a target system. Also shown is the estimated 
deposited power.}
\label{fig:powerGeometry}
\end{figure}

Figure~\ref{fig:powerGeometry} shows that the power from the 4\,MW proton beam
is distributed throughout the whole target station volume, with more than half
of the total power deposited in the water-cooled tungsten-carbide (WC) shielding.
About 10\% of the total power is deposited in the mercury jet, with
high energy deposition also seen in the normal conductor magnet coils. The major
concern is that about 53\,kW of power is absorbed by the first superconducting coil
surrounding the target (SC1), as well as significant power deposition observed
for some of the other coils. This heat load is extremely high, with an estimated 
peak power of 70\,mW/cc (10\,mW/g) for SC1, which is beyond the
design guidelines for a maximum superconducting heat load less than 1\,mW/cc 
for the International Thermonuclear Experimental Reactor (ITER)~\cite{ref:ITER}.
The amount of heat deposited in the superconducting coils increases when the
tungsten-carbide shielding (80\% WC and 20\% water) is replaced with 
either pure tungsten, mercury or a combination of the two
(see table~\ref{tab:standardGeometry}). Increasing the shielding volume 
from an outer radius $r$ of 50\,cm to 63\,cm, up to the SC1 inner boundary (enhanced
shielding option), helps to reduce the heat loads in the coils by approximately 
a factor of three, but this is still inadequate.
\begin{table}[htb!]
\caption{Estimated deposited power (kW) in the Study 2a 
superconducting coils and shielding for various shielding materials using 
FLUKA simulations. The percentages denote the relative amount by weight of each
material (uniformly distributed within the shielding volume).}
\begin{center}
\begin{tabular}{|cccc|}
\hline
Shielding Material & SC1 & SC1-SC13 & Shielding \\
\hline
80\% WC + 20\% Water & 54 & 64  & 2293\\
100\% W              & 57 & 70  & 2231\\
80\% W + 20\% Hg     & 63 & 80  & 2199\\
60\% W + 40\% Hg     & 69 & 91  & 2171\\
100\% Hg             & 87 & 134 & 2083\\
\hline
\end{tabular}
\end{center}
\label{tab:standardGeometry}
\end{table}

The heat load estimates shown above are consistent with those found using
MARS simulations~\cite{ref:MARSEDep}, except for the superconducting coils, 
which are a factor of two higher. Investigations of the energy deposition in 
simple target geometries show that the particle (hadronic) showers in FLUKA 
are more penetrating than those in MARS. Further studies show that much
better agreement is achieved when the MCNP mode~\cite{ref:MCNP} for MARS is enabled,
giving a consistent treatment of the transport of very low energy (thermal)
neutrons, with a kinetic energy cut-off of $10^{-14}$\,GeV,
that can pass through the shielding and deposit energy in the superconducting coils.

\begin{figure}[hbt]
\centering
{\includegraphics*[width=0.4\textwidth,bb=0 0 520 390]{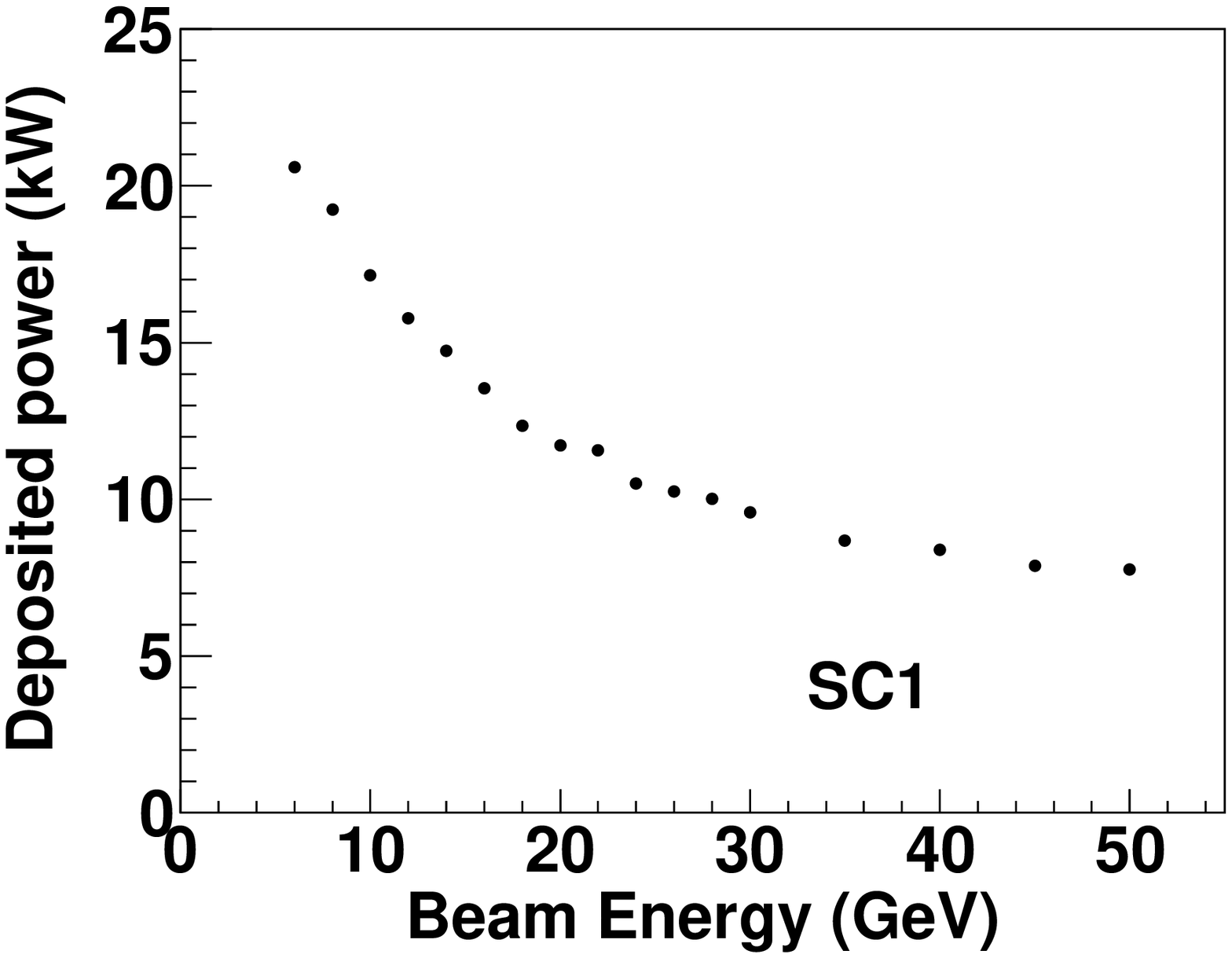}
\includegraphics*[width=0.4\textwidth,bb=0 0 520 390]{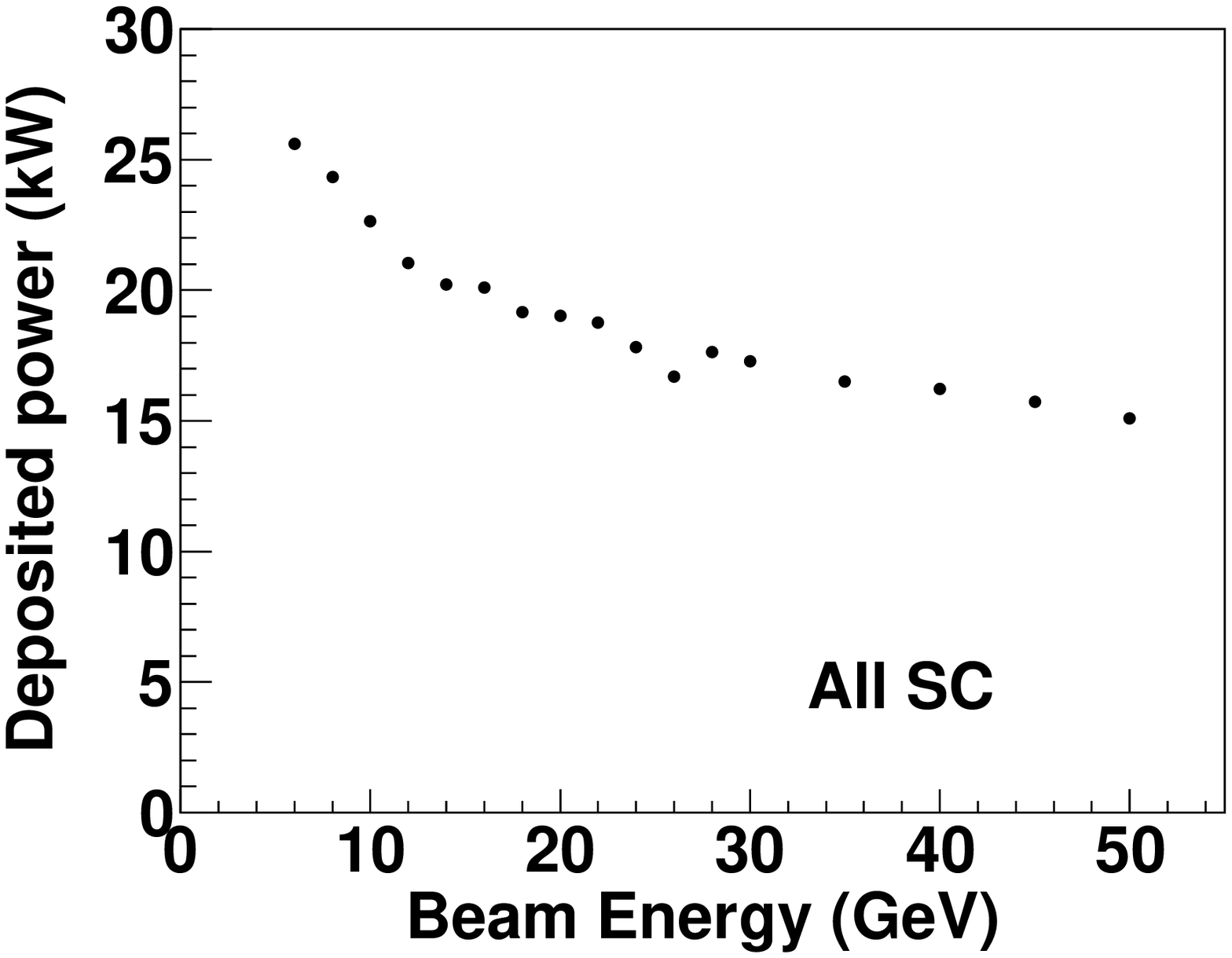}}
{\includegraphics*[width=0.4\textwidth,bb=0 0 520 390]{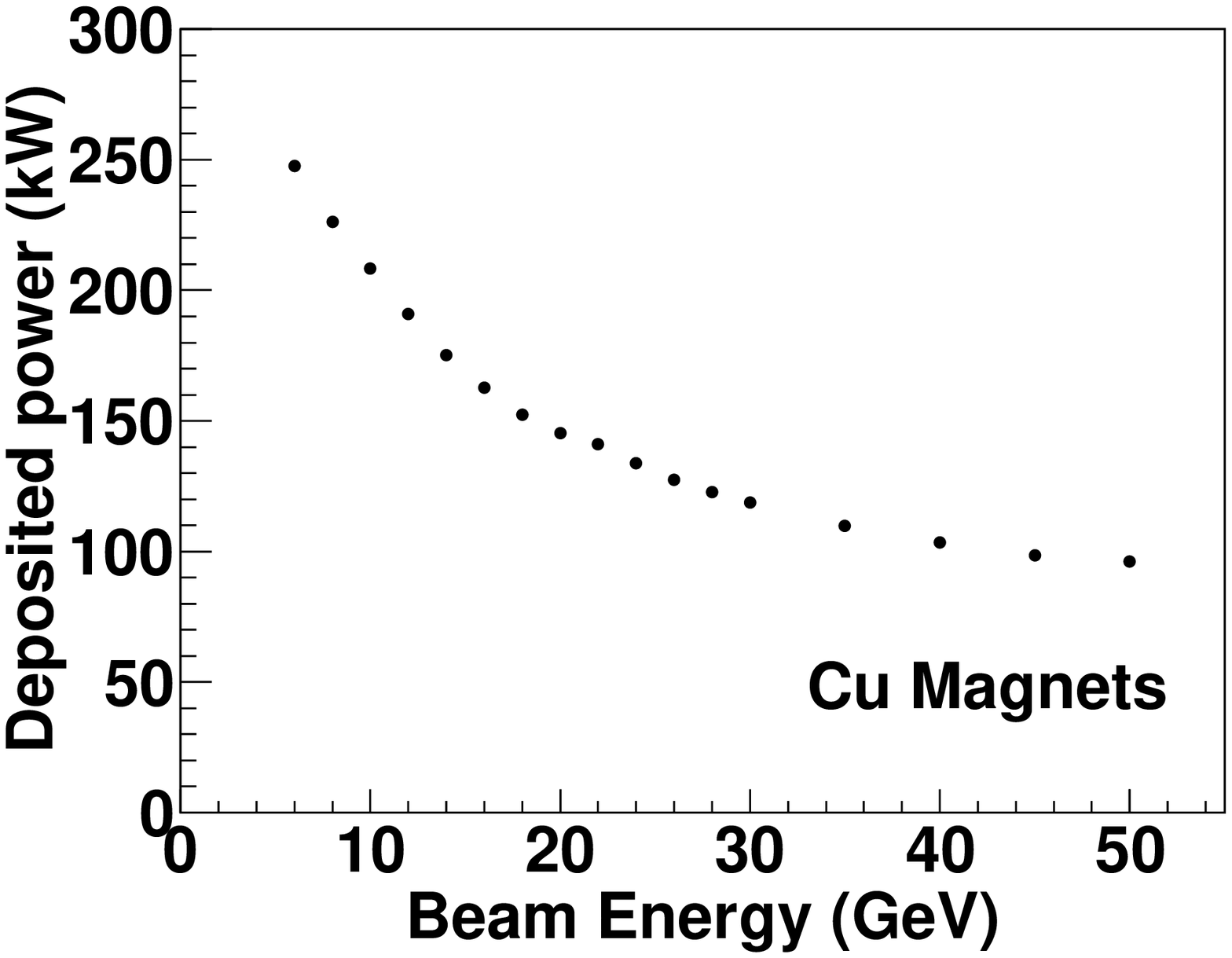}
\includegraphics*[width=0.4\textwidth,bb=0 0 520 390]{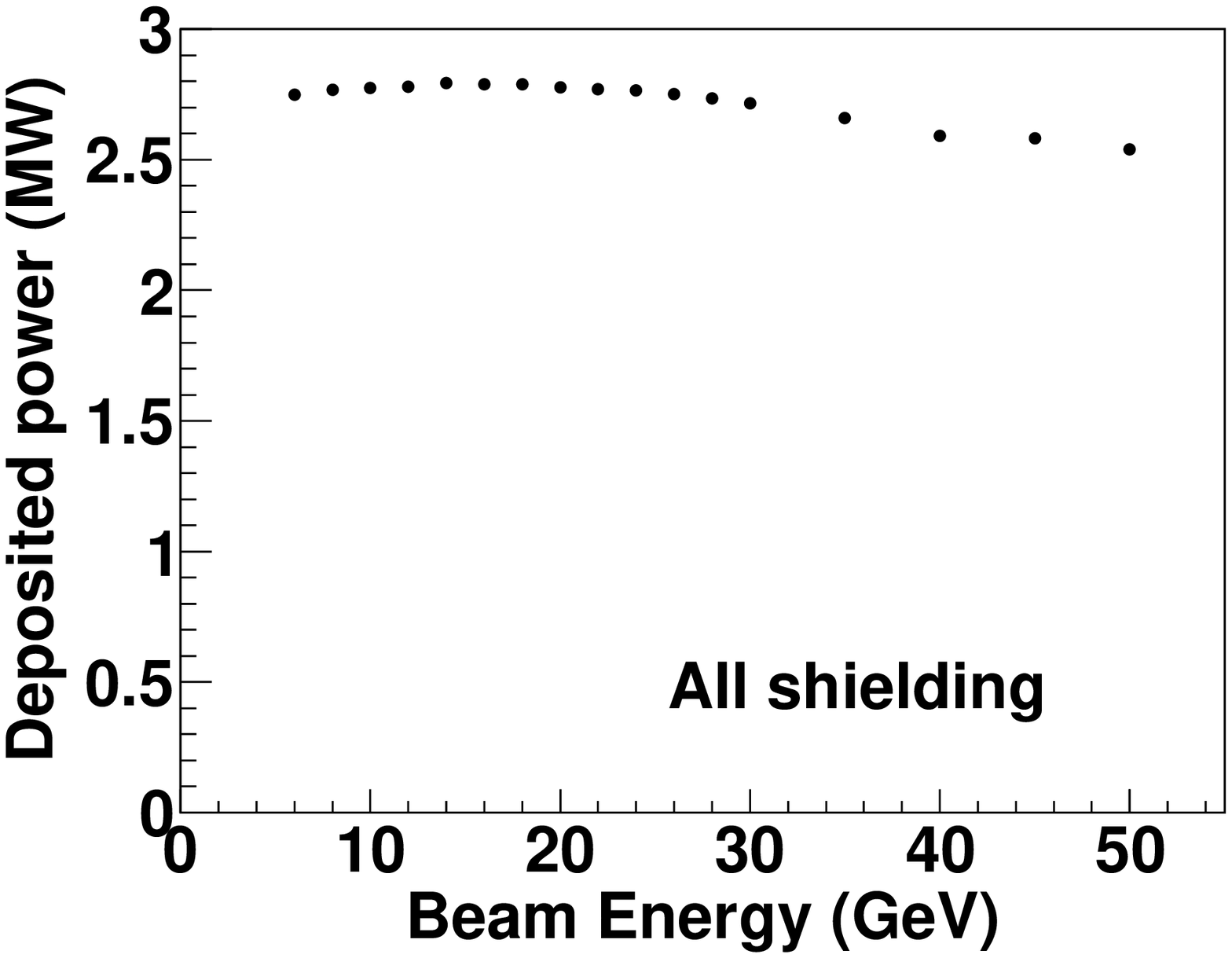}}
{\includegraphics*[width=0.4\textwidth,bb=0 0 520 390]{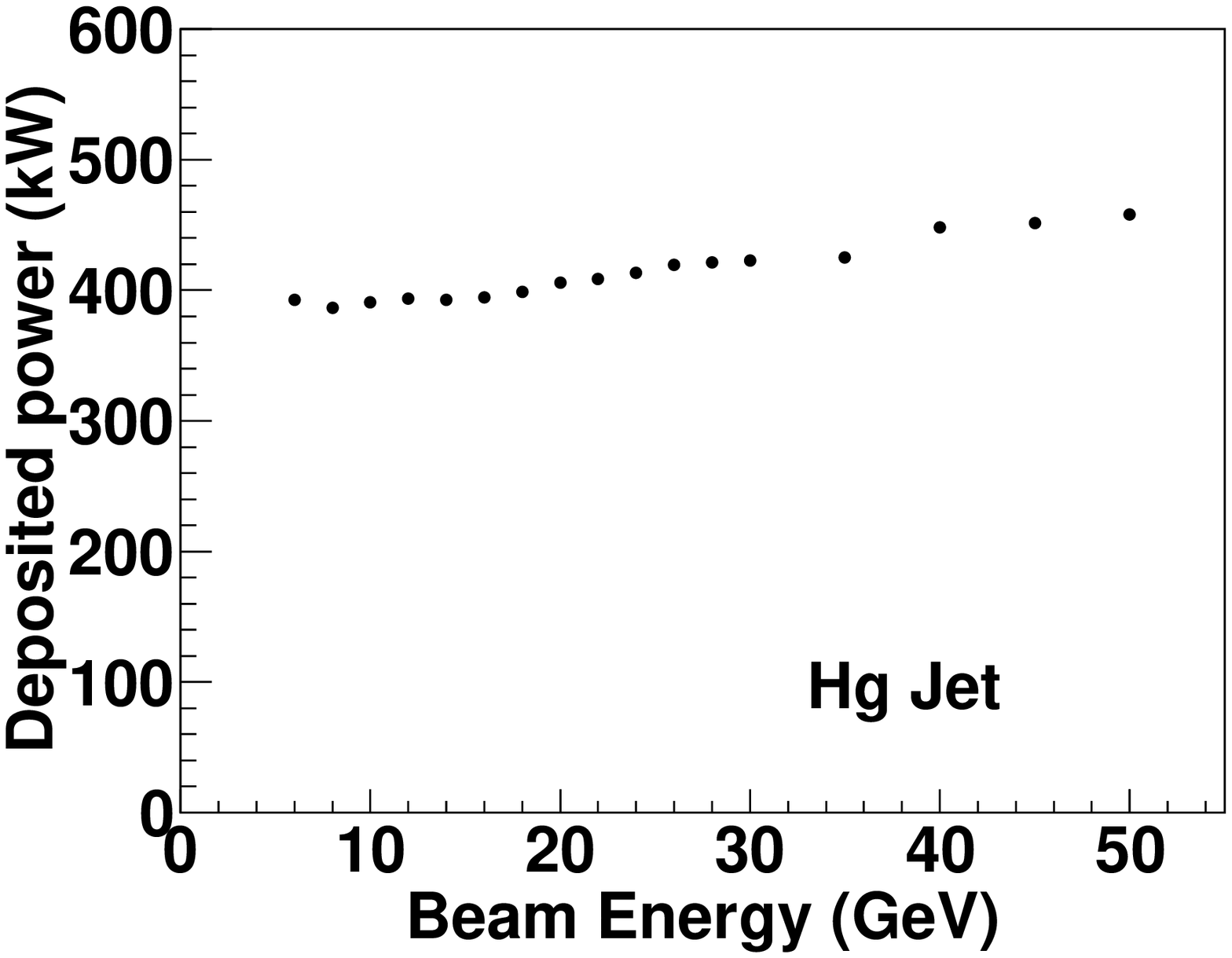}
\includegraphics*[width=0.4\textwidth,bb=0 0 520 390]{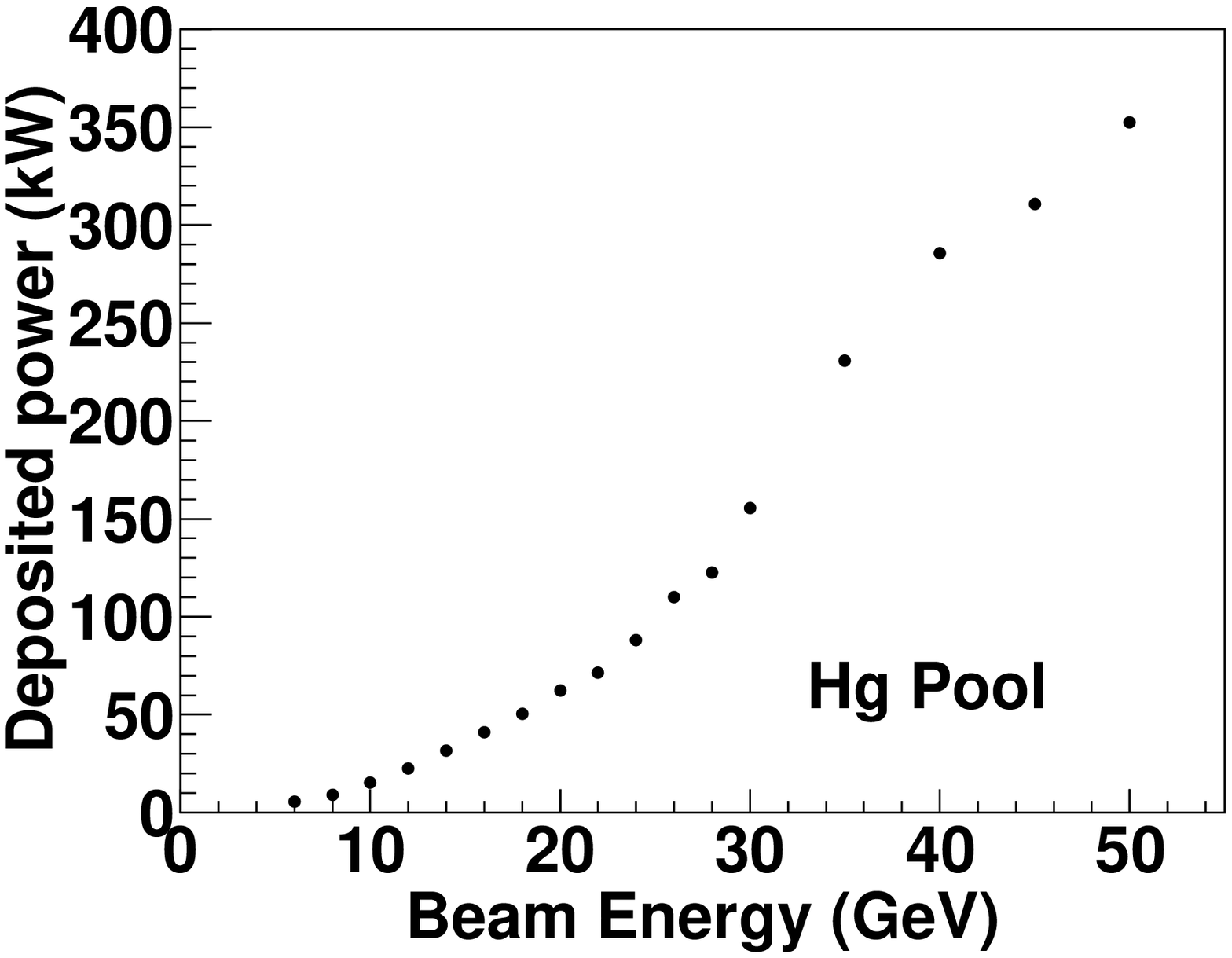}}
\caption{Variation of the average total deposited power in
different parts of the Study 2a target system, with enhanced shielding, 
as a function of the input proton beam kinetic energy (for 4\,MW).}
\label{fig:PowervsBeamEnergy}
\end{figure}
The variation of the deposited power in several regions
of the target station, with enhanced shielding up to a radius of 63\,cm,
as a function of the initial proton beam kinetic energy $E_{\rm{beam}}$ is 
shown in figure~\ref{fig:PowervsBeamEnergy}, all for a total beam power of 
4\,MW (the proton rate is inversely proportional to the input kinetic energy). 
Beam energies below 5\,GeV are not considered here,
since FLUKA uses another hadron-nucleon model to simulate the beam-target
interaction that differs to the one used at higher energies.
For the magnet coils, the deposited power is reduced as the 
proton beam energy increases. The overall power absorbed by the shielding stays roughly 
constant, and there is only a slight increase in the deposited power in the mercury jet
at higher beam energies. In contrast, the heat dissipation from
secondary particle interactions in the mercury pool dramatically increases 
with input energy. Note that the energy of the mercury jet will also be absorbed 
by the mercury pool as the jet flows into it, so the effective energy deposition 
for the mercury flow system is the sum of the separate jet and pool contributions.
From these results, it is obvious that the problems of significant
energy deposition in the superconducting coils can not be solved
by just changing the beam energy, slightly adjusting the shielding volume or 
modifying the material used in the shielding.

\section{Increased shielding geometry}

\begin{figure}[htb]
\centering
\includegraphics*[width=0.8\textwidth]{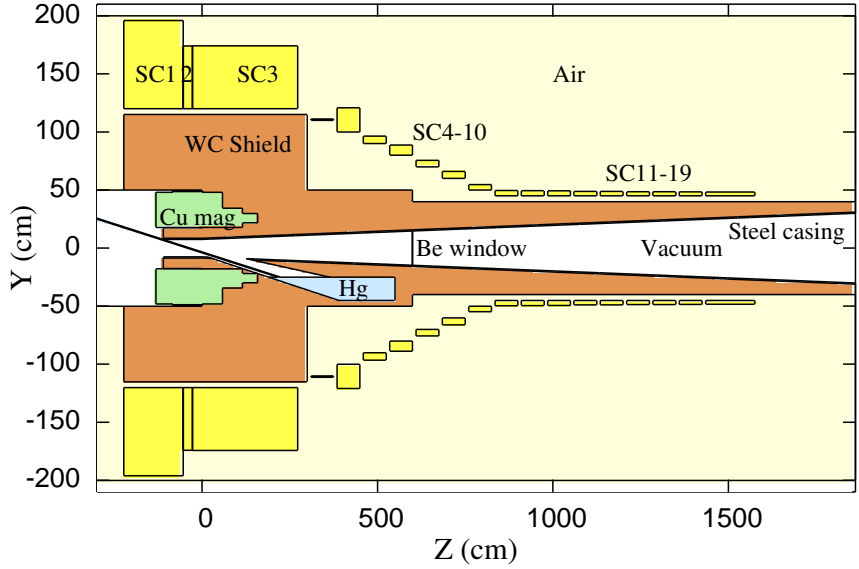}
\caption{New target geometry incorporating a substantial increase to the shielding 
volume inside the superconducting coils (SC). The materials used for SC1--9 and SC10--19 
are Nb$_{3}$Sn and NbTi, respectively.}
\label{fig:IDS120fGeometry}
\end{figure}

The results shown earlier demonstrate that the
original Study 2a geometry for the Neutrino Factory target station
suffers from a number of problems regarding energy deposition. 
Specifically, the first few superconducting coils experience a radiation dose 
that is too high for their safe operation. We will show that this can been mitigated
by effectively doubling the outer radius of the shielding protecting the inside of
the superconducting magnets.

\begin{figure}[hbt]
\centering
\includegraphics*[width=0.8\textwidth]{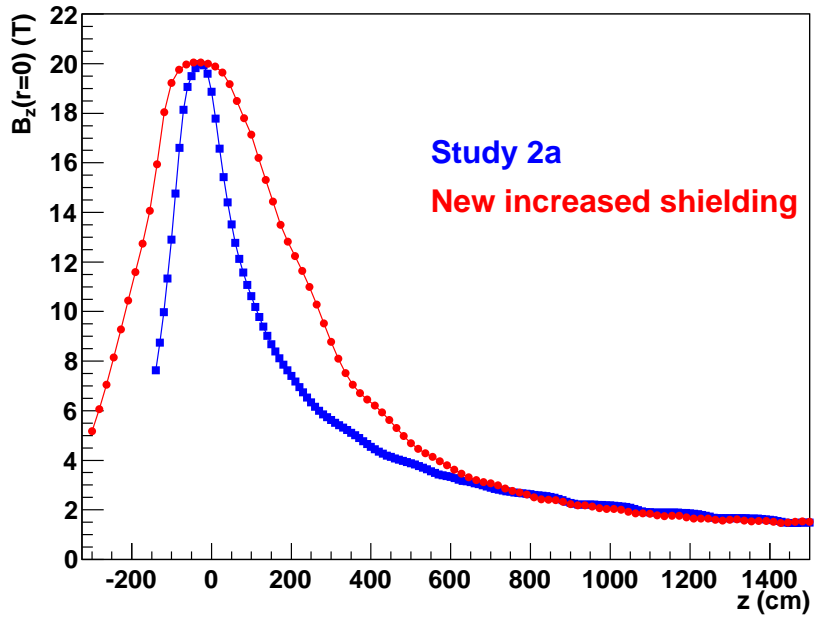}
\caption{Comparison of the $B_z$ magnetic field profile along the $z$ axis for the Study 2a
(blue squares) and the new increased shielding (red circles) geometries.}
\label{fig:BzFieldComparison}
\end{figure}

Figure~\ref{fig:IDS120fGeometry} shows a schematic of a 
new target station geometry which contains substantially increased shielding to limit
the heat deposition in the coils, whose inner radii have been increased
from 63\,cm to 120\,cm. Note that the layout of all of the coils
has been changed to ensure that the magnetic field at the
beam-target interaction region ($z=-37.5$\,cm) peaks at 20\,T.
The parameters defining the new coil positions and current densities, as a result
of the work presented in~\cite{ref:RWeggel}, are shown in table~\ref{tab:IDS120fMagParams}.
A comparison of the magnetic field along the $z$ axis, $B_z(r=0)$,
between the Study 2a and increased shielding geometries is shown in 
figure~\ref{fig:BzFieldComparison}. For the new geometry, the magnetic 
field distribution has a broader peak, but has a field taper for $z$ greater than 600\,cm
that closely matches the taper from the Study 2a geometry.
The tungsten-carbide shielding
contains a larger fraction of cooling water (40\%), while the iron plug just behind
the normal conductor coils has been removed for simplicity. The same 8\,GeV proton beam
and mercury jet parameters are used as before.

\begin{table}[thb!]
\caption{Parameters defining the normal conducting (NC) and superconducting
(SC) coils for the new increased shielding geometry shown in figure 4: $z_0$ is the initial $z$ position, 
$\Delta z$ specifies the length along $z$, $r_1$ is the inner radius, $\Delta r$ is the radial thickness,
while $I$ denotes the average current density. The last column specifies the coil materials used in
the FLUKA simulation.}
\begin{center}
\begin{tabular}{|lllllll|}
\hline
Label & $z_0$ (cm) & $\Delta z$ (cm) & $r_1$ (cm) & $\Delta r$ (cm) & $I$ (A/mm$^2$) & Material \\
\hline
NC1   & -131.3     & 47.3            & 17.8       & 30.2           &  16.6          & Cu \\
NC2   &  -84.0     & 86.2            & 17.8       & 30.9           &  16.6          & Cu \\
NC3   &    2.1     & 56.2            & 17.8       & 30.3           &  16.6          & Cu \\
NC4   &   58.3     & 57.0            & 17.8       & 16.6           &  16.6          & Cu \\
NC5   &  115.3     & 43.5            & 21.9       &  8.0           &  16.6          & Cu \\
SC1   & -222.6     & 169.4           & 120.0      & 75.9           &  23.2          & Nb$_3$Sn \\
SC2   &  -53.1     & 26.1            & 120.0      & 54.0           &  23.2          & Nb$_3$Sn \\
SC3   &  -27.1     & 300.0           & 120.0      & 54.1           &  23.1          & Nb$_3$Sn \\
SC4   &  310.0     & 65.0            & 110.0      &  1.2           &  30.0          & Nb$_3$Sn \\
SC5   &  385.0     & 65.0            & 100.0      & 20.8           &  33.3          & Nb$_3$Sn \\
SC6   &  460.0     & 65.0            & 90.0       &  6.4           &  35.9          & Nb$_3$Sn \\
SC7   &  535.0     & 65.0            & 80.0       &  8.7           &  38.2          & Nb$_3$Sn \\
SC8   &  610.0     & 65.0            & 70.0       &  5.6           &  40.0          & Nb$_3$Sn \\
SC9   &  685.0     & 65.0            & 60.0       &  6.1           &  40.0          & Nb$_3$Sn \\
SC10  &  760.0     & 65.0            & 50.0       &  4.7           &  40.0          & NbTi \\
SC11  &  835.0     & 65.0            & 45.0       &  4.6           &  40.0          & NbTi \\
SC12  &  910.0     & 65.0            & 45.0       &  4.4           &  40.0          & NbTi \\
SC13  &  985.0     & 65.0            & 45.0       &  4.3           &  40.0          & NbTi \\
SC14  & 1060.0     & 65.0            & 45.0       &  3.9           &  40.0          & NbTi \\
SC15  & 1135.0     & 65.0            & 45.0       &  3.8           &  40.0          & NbTi \\
SC16  & 1210.0     & 65.0            & 45.0       &  3.5           &  40.0          & NbTi \\
SC17  & 1285.0     & 65.0            & 45.0       &  3.5           &  40.0          & NbTi \\
SC18  & 1360.0     & 65.0            & 45.0       &  3.4           &  40.0          & NbTi \\
SC19  & 1435.0     & 140.0           & 45.0       &  3.2           &  40.0          & NbTi \\
\hline
\end{tabular}
\end{center}
\label{tab:IDS120fMagParams}
\end{table}
\begin{figure}[htb]
\centering
\includegraphics*[width=0.85\textwidth]{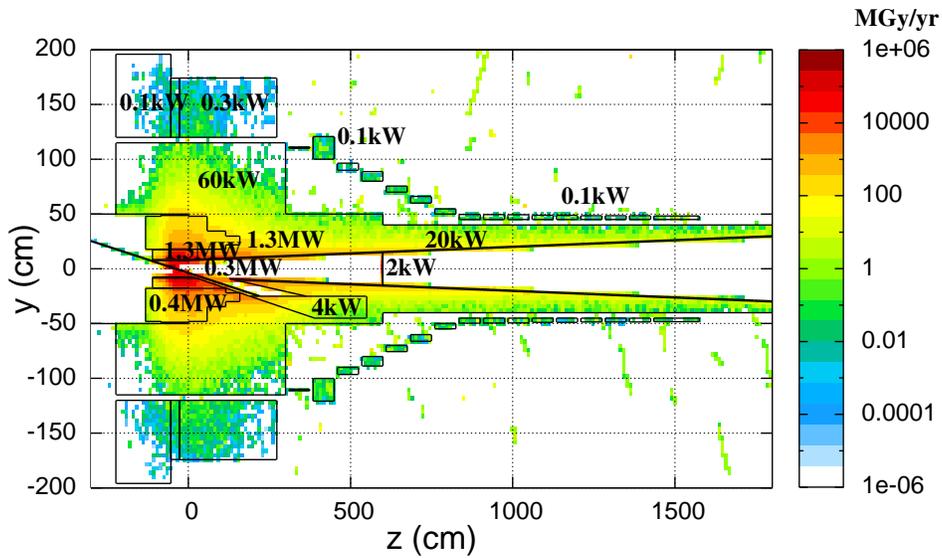}
\caption{Distribution of the total radiation dose ($10^{6}$ Gy/year) in
the new target station geometry. Also shown is the estimated deposited power.}
\label{fig:IDS120fPower}
\end{figure}
\begin{table}[thb!]
\caption{Estimated deposited power in various regions of the new target
station geometry shown in figures 4 and 6. The labels SC and NC denote the superconducting and 
normal conducting magnets, respectively.}
\begin{center}
\begin{tabular}{|ll|}
\hline
Region & Power (kW) \\
\hline
SC1 & $0.05 \pm 0.01$ \\
SC2 & $0.03 \pm 0.01$ \\
SC3 & $0.26 \pm 0.06$ \\
SC4 & $<0.01$ \\
SC5 & $0.07 \pm 0.01$ \\
SC6 to SC10 & $0.08 \pm 0.01$\\
SC11 to SC19 & $0.07 \pm 0.01$\\
Inner shielding $r<17.5$\,cm, $z<0$\,m & $1050 \pm 10$\\
Shielding $0 < z < 6$\,m, $r < 50$\,cm & $1018 \pm 10$\\
Downstream shielding $z > 6$\,m & $20 \pm 1$\\
SC shield $r > 50$\,cm & $61 \pm 2$\\
Inner shield casing $z < 0$\,m & $238 \pm 5$\\
Inner shield casing $z > 0$\,m & $245 \pm 5$ \\
Hg jet & $319 \pm 5$\\
Hg pool & $4.4 \pm 0.4$\\
NC1 & $2.1 \pm 0.2$\\
NC2 & $188 \pm 5$\\
NC3 & $140 \pm 4$\\
NC4 & $59 \pm 3$\\
NC5 & $16 \pm 2$\\
4\,mm Be window at 6\,m & $2.1 \pm 0.2$ \\
\hline
\end{tabular}
\end{center}
\label{tab:IDS120fPower}
\end{table}
\begin{figure}[hbt]
\centering
\includegraphics*[width=0.8\textwidth]{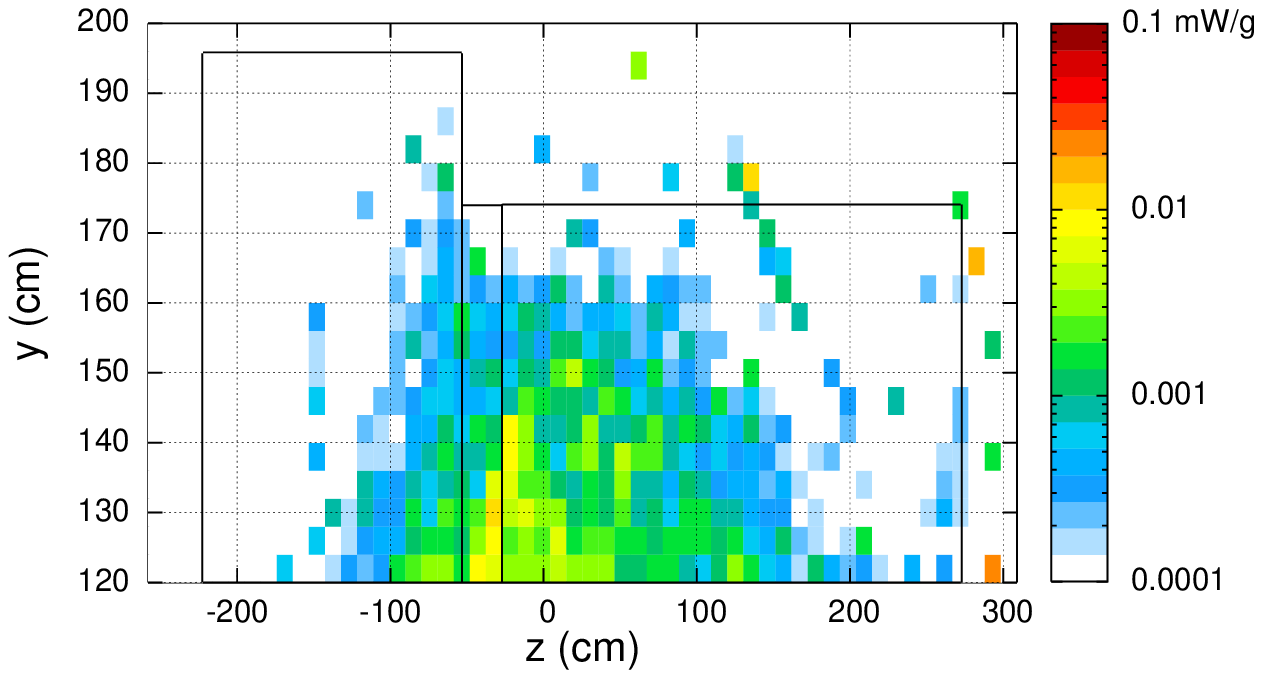}
\includegraphics*[width=0.8\textwidth]{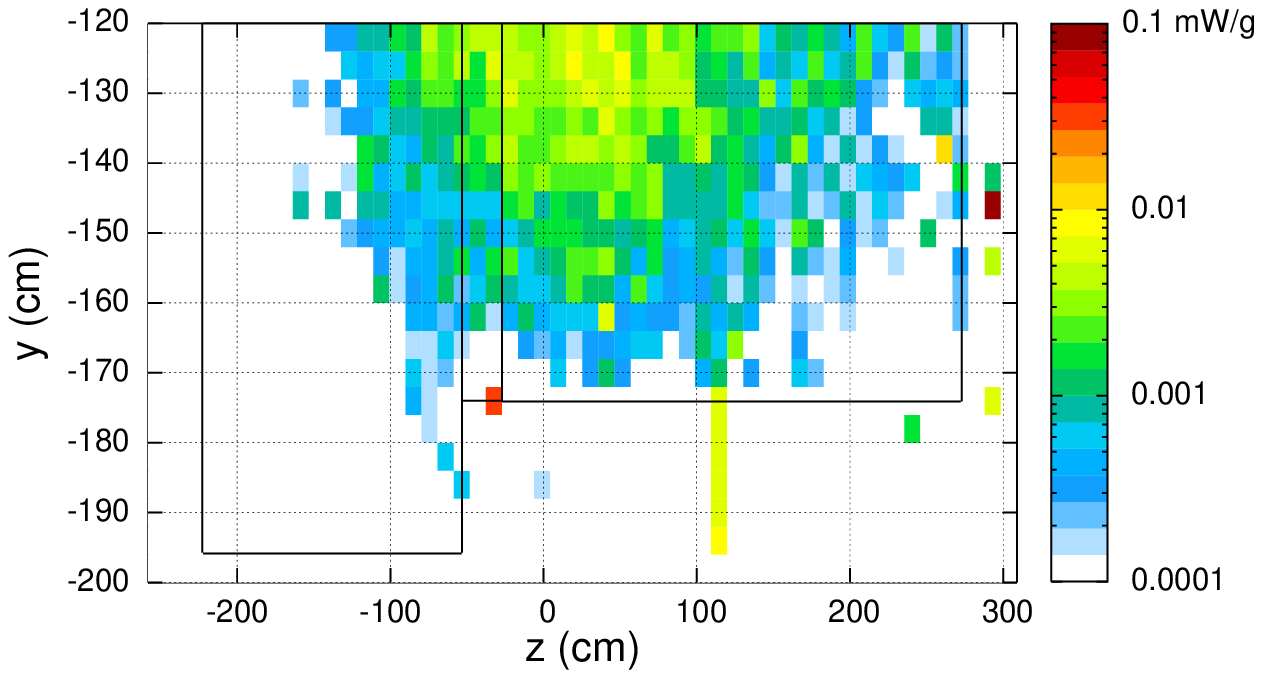}
\caption{Distribution of the specific deposited power (mW/g) in the first three
superconducting coils above (top: $y>0$) and below (bottom: $y<0$) the $z$ axis.}
\label{fig:SC3Power}
\end{figure}

Figure~\ref{fig:IDS120fPower} shows the estimated total radiation dose 
in units of $10^6$ Gy for $2\times 10^7$\,s/year
(to be conservative), while table~\ref{tab:IDS120fPower} 
shows the deposited power in various regions of the new target station,
where the uncertainties are estimated by using different initial random number
seeds for the FLUKA simulation. We can immediately see that the power depositions
in the superconducting coils have all been reduced to values below 1\,kW, with
the maximum value around 0.3\,kW occurring for the third coil. The extended shielding 
for $r > 50$\,cm stops $\sim$60\,kW of heat from reaching the first few 
superconducting coils.
The combined shielding still absorbs just over half of the total beam power, 
with around 8\% deposited in the mercury jet. However, the heat loads in the normal
conducting copper coils located inside the shielding volume, whose lengths have 
approximately doubled, have increased by a factor of two.

Figure~\ref{fig:SC3Power} shows a close-up view of the deposited power per unit mass
for the first three superconducting coils (SC1--3). Slightly more energy is deposited
in the $y<0$ region since the proton beam is pointing down in this direction.
The peak energy density for SC3
is less than 0.05\,mW/g, which is below the ITER requirement of 
0.17\,mW/g~\cite{ref:ITER}, assuming a conductor mass density of 6\,g/cc. 
This peak energy deposition is equivalent to $10^6$\,Gy/year 
(for a year of $2\times 10^7$\,s), which is well below the estimated maximum 
allowed integrated dose of $10^8$\,Gy~\cite{ref:Study2a}, meaning that the coils
should have sufficient operational lifetime. These results are consistent
with those obtained using MARS with the MCNP mode enabled~\cite{ref:NSouchlas}.

A major concern with the new target station design is that there will be
a four-fold increase in the stored energy of the magnets
approaching 1\,GJ. This energy needs to be managed safely in the event of a quench,
in which a section of the magnet becomes too hot, losing its superconductivity 
and affecting other nearby coils. The larger radius of the coils also means 
that the magnetic forces ($\sim1$\,k$-10$\,kTonnes) between them increases, causing 
difficulties for the overall magnet support structure~\cite{ref:MagForceStudy}.
To help with the significant engineering challenges, it may be necessary
to reduce the achievable magnetic field in the target interaction region,
involving a trade-off between magnet construction feasibility and an acceptable 
loss of useful low-energy pions from the target.
\section{Radiation damage}

\begin{figure}[htb]
\centering
\includegraphics*[width=0.8\textwidth]{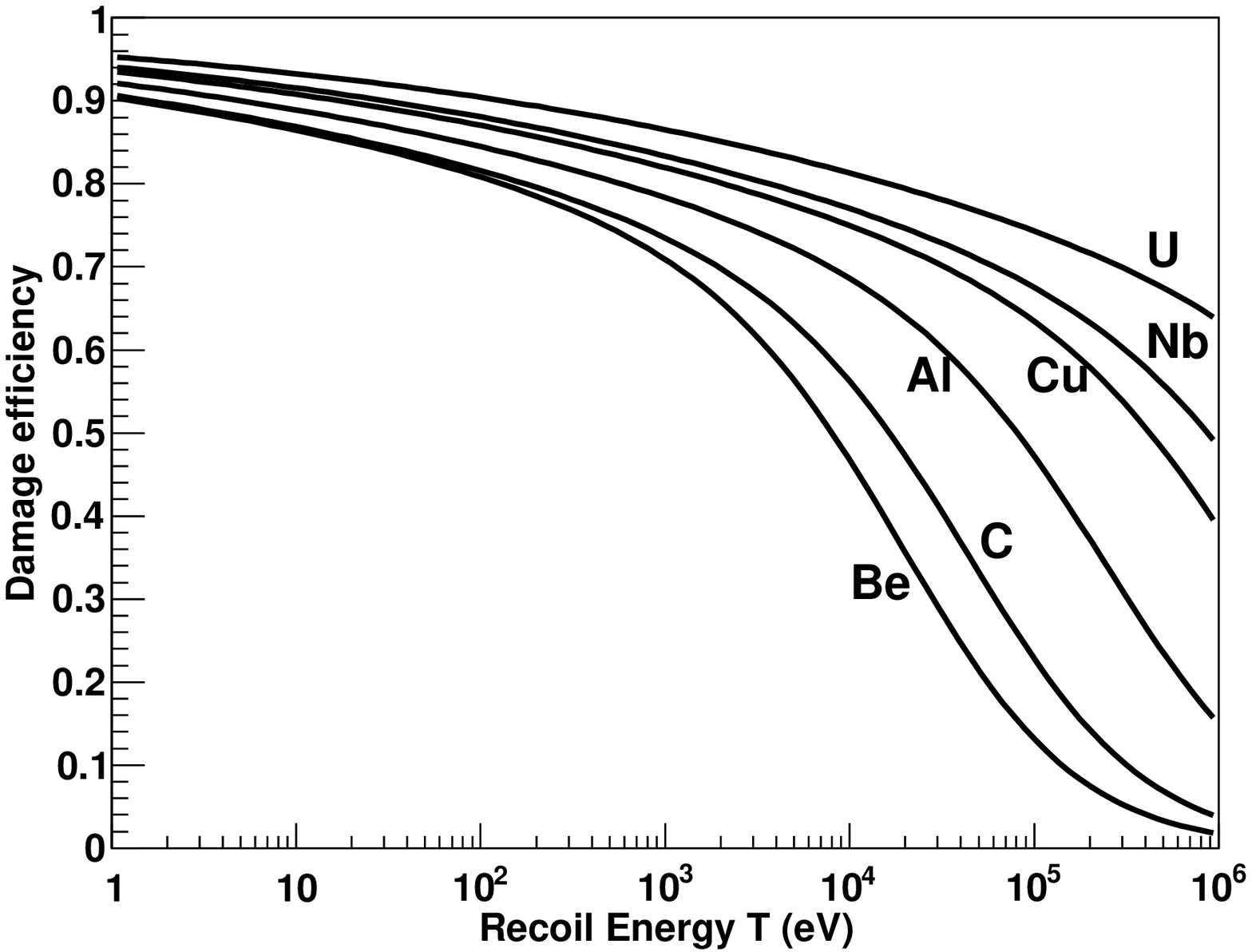}
\caption{Distribution of the damage efficiency function $\xi(T)$ for several materials
as a function of the recoil (kinetic) energy of the primary knock-on atom.}
\label{fig:PartitionFun}
\end{figure}
The target station will be a high radiation environment, and it
is important to know how this will affect various materials used in its
construction. For instance, radiation damage to the superconducting
coils will severely limit their operational capability, and may even
make them inoperable. One measure of this damage
is the average number of displacements per atom (DPA) inside a material.
For example, a radiation dose of 2 DPA means that each atom in the material 
has been displaced within the structural lattice an average of two times.
Here, the DPA values for several regions of the new
target station geometry are estimated using an empirical model based
on the LSS theory of atomic collisions in solids~\cite{ref:LSSTheory,ref:DPAModel}.
In this model, a (Frenkel) defect forms when an atom is given energy
and leaves its place in the atomic lattice, 
displacing nearby atoms in a cascade effect. The initial displaced 
atom is known as the primary knock-on atom (PKA). The number of defects 
(per incident proton) is given by
\begin{equation}
N_F = \kappa \frac{\xi(T)T}{2E_{th}},
\label{eqn:N_F}
\end{equation}
where $T$ is the kinetic energy of the PKA, which we assume is equal to the amount of
deposited energy, $\xi(T)$ is the partition function describing how the initial 
energy displaces other atoms within the cascade, $E_{th}$ is the damage threshold energy, 
which is the minimum energy needed to displace an atom within the lattice, and 
$\kappa \approx 0.8$ is the displacement efficiency. 
Example values of $E_{th}$ are 10\,eV for beryllium,
40\,eV for copper and niobium, and 90\,eV for tungsten. The factor of two
in the denominator is due to the fact that the model assumes
that the energy in each two-body collision within the cascade is shared equally. 
Figure~\ref{fig:PartitionFun} shows the distribution of the partition function 
$\xi(T)$, also known as the damage efficiency, for several materials. 
In general, heavier atomic mass elements have larger
$\xi(T)$ values for a given kinetic energy $T$.
The total number of DPA per year is then given by
\begin{equation}
{\rm{DPA/year}} = \langle N_F \rangle \frac{N_p}{\rho V},
\label{eqn:DPA}
\end{equation}
where $\langle N_F \rangle$ is the average value of $N_F$ per proton, $\rho$ is the atomic
density, $V$ is the volume of the given material region, and $N_p$
is the total number of protons on target per year ($2\times10^7$\,s), 
which is equal to $6.25 \times 10^{22}$/year for a 4\,MW, 8\,GeV proton beam.

Using the previous energy deposition results for the new target geometry, we 
find that an upper limit of the radiation dose for the 
superconducting magnets (assuming Nb-Nb atomic collisions dominate) is approximately
$3 \times 10^{-4}$\,DPA/year. This is about a factor of ten lower than 
the critical operational dose of $1.9 \times 10^{-3}$ for radiation (neutron)
damage that can cause irreversible reductions of allowed currents in
Nb$_3$Sn superconducting coils~\cite{ref:Nb3SnDPA}. However, there will need to be 
ceramic insulating shielding for the magnet conductors, and it is not known
at present how they may be damaged by the radiation dose.
For comparison, the radiation dose for the first superconducting coil
(SC1) in the Study 2a geometry is estimated to be $2 \times 10^{-3}$\,DPA/year, 
which is equal to the critical dose.

The peak radiation doses for the normal conducting magnets and the shielding 
in the new geometry are approximately 0.2\,DPA/year and 1.2\,DPA/year, respectively.
The steel casing along the inside of the shielding (``beam pipe'') suffers a 
radiation dose of $\sim 5$\,DPA/year close to the interaction region, which lowers
to a value below 0.1\,DPA/year further downstream.

A potential concern is that a radiation dose of 0.9\,DPA/year
is estimated for the 4\,mm beryllium window, which is located at $z=6$\,m
to stop mercury (liquid and vapour) and rare gas reaction products
from going further downstream. Such a dose may deform the window, and there
will also be significant production of tritium, which 
has a half-life around 12 years and is a potential radiation hazard. This
means that the window will need to be regularly replaced in order
to protect the downstream components of the Neutrino Factory accelerator
system.

The activation of the mercury jet system
will be dominated by the production of mercury isotopes, although there will also
be significant isotope production of nearby elements such as gold, platinum
and iridium, all of which typically decay within a few hours or days.
The large number of different isotopes in the mercury jet system may cause
material compatibility concerns for the mercury return flow and nearby 
cooling loops.

\section{Failure modes}

It is important to know how the beam power will be distributed within the
target station for a range of failure modes, such as if there is a complete failure 
of the magnetic field ($\underline{B}$) or the mercury jet stops flowing and the 
full 4\,MW proton beam goes straight into the mercury pool reservoir 
and shielding beam dump. Table~\ref{tab:OtherCases} shows comparisons of the 
average power deposition in various regions of the new target station geometry 
between normal and extreme operating conditions.
\begin{table}[hbt!]
\setlength\tabcolsep{8pt}
\caption{Average power deposition (kW) for the new target geometry 
(see figure 4) under normal and extreme operating scenarios. The labels SC and NC denote the superconducting and 
normal conducting magnets, respectively.}
\begin{tabular}{|lllll|}
\hline
Region & Normal & No $\underline{B}$ & No Hg jet & No $\underline{B}$ \& No Hg jet \\
\hline
SC1               & $0.05 \pm 0.01$ & $0.04 \pm 0.01$ & $<0.01$ & $<0.01$ \\
SC2               & $0.03 \pm 0.01$ & $0.03 \pm 0.01$ & $<0.01$ & $<0.01$ \\
SC3               & $0.26 \pm 0.06$ & $0.28 \pm 0.06$ & $0.22 \pm 0.05$ & $0.26 \pm 0.06$ \\
SC4               & $<0.01$ & $0.35 \pm 0.06$ & $0.01 \pm 0.01$ & $1.1 \pm 0.1$ \\
SC5               & $0.07 \pm 0.01$ & $3.3 \pm 0.3$ & $0.09 \pm 0.01$ & $9.2 \pm 1.0$ \\
SC6 to SC10       & $0.08 \pm 0.01$ & $0.47 \pm 0.08$ & $0.05 \pm 0.01$ & $1.0 \pm 0.1$ \\
SC11 to SC19      & $0.07 \pm 0.01$ & $0.02 \pm 0.01$ & $0.02 \pm 0.01$ & $<0.01$ \\
Shielding           & $2149 \pm 14$ & $2096 \pm 14$ & $2567 \pm 20$ & $1693 \pm 12$ \\
Inner shield casing & $483 \pm 7$ & $301 \pm 6$ & $627 \pm 9$& $17 \pm 1$ \\
Hg jet              & $319 \pm 5$ & $183 \pm 3$ & --- & --- \\
Hg pool             & $4.4 \pm 0.4$ & $640 \pm 10$ & $13 \pm 1$ & $1835 \pm 12$ \\
NC1 to NC5          & $405 \pm 7$ & $329 \pm 6$ & $350 \pm 6$ & $4.0 \pm 0.5$ \\
4\,mm Be window     & $2.1 \pm 0.2$ & $0.02 \pm 0.01$ & $0.01 \pm 0.01$ & $<0.01$ \\
\hline
\end{tabular}
\label{tab:OtherCases}
\end{table}

With no magnetic field present to steer the proton beam (and any secondary
charged particles), the energy deposition for the combined mercury jet and pool 
system increases by a factor of $\sim 2.5$.
In addition, the downstream superconducting coils
see a large increase in radiation dose, but this can
be reduced to safe levels by introducing an additional conic frustum to the
shielding for $z=3$ to 6\,m, with a radius between 50 and 100\,cm. Note that the
power deposited in the mercury jet itself decreases, owing to the fact
that the proton beam, which is no longer bending in the magnetic field,
intersects a smaller mercury target volume. If no mercury jet is flowing,
then most of the beam power ($\sim$80\%) is dumped in the 
water-cooled tungsten-carbide shielding and inner beam pipe steel casing. The
presence of the magnetic field and the geometry of the surrounding region
ensures that little energy is deposited in the mercury pool reservoir.
In contrast, with no magnetic field and no mercury jet present, the mercury pool
experiences a large thermal shock equivalent to just below half of the
total beam power. This will produce significant agitation of the mercury
pool surface with splashes of radial velocities expected to approach
50\,m\,s$^{-1}$~\cite{ref:HgSplash}.

Concerning the calculation of DPA values, it will be advantageous 
to quantify the radiation doses within much
smaller regions of the target station geometry in order
to identify local hot-spots of radiation damage. In addition, an improved model
for calculating DPA values to much better accuracy is required.
During the time of this study, the FLUKA simulation package
did not have this capability, but such a feature is expected to be made
available in the near future.

\section{Summary}

We have shown that a lot of shielding is required
to protect the superconducting coils from large radiation doses
in the Neutrino Factory target station, as provided by the current versions
of the FLUKA and MARS simulation codes. This, however, poses additional engineering 
challenges for the magnet support design. The shielding absorbs about half of the 
total beam power, which will have to be handled by any water cooling system. 
There will be significant energy deposition
for the normal conducting magnets, and the beam pipe casing close
to the interaction region will experience large radiation doses.
There may also be safety issues with the radiation dose expected
for the beryllium window protecting the rest of the decay channel
from mercury liquid/vapour. 
The energy deposition results are essentially unchanged if 
the mercury jet is assumed to only have an effective length of 30\,cm,
equivalent to approximately two interaction lengths. Finally, we should expect
to observe similar levels of radiation dose throughout the whole target
station if we replace the mercury jet with a solid or powdered jet
target made of tungsten.

\section{Acknowledgements}

We acknowledge the financial support of the European Community under
the European Commission Framework Programme 7 Design Study: EUROnu,
Project Number 212372. The EC is not liable for any use that may
be made of the information contained herein. I also
thank my colleagues from the International Design Study (IDS-NF)
collaboration for fruitful discussions concerning this work.

\end{document}